
\input phyzzx.tex
\def\neqn#1{\eqn#1}   
\def\E#1{#1}
%
%
\def\Rc#1{[#1]}
\def\tytul#1{}         
\def\name#1{}
%
\catcode`\@=11 
\def\refitem#1{\r@fitem{[#1]}}
\def\myjournal#1&#2(#3){\begingroup \let\myjournal=\dummyj@urnal
     #1\unskip~\ignorespaces #2
    (\afterassignment\j@ur \count255=#3) \endgroup\ignorespaces }
\def\refout{\par\penalty-400\vskip\chapterskip
   \spacecheck\referenceminspace
   \ifreferenceopen \Closeout\referencewrite \referenceopenfalse \fi
   \line{{\bf References}\hfil}\vskip\headskip
   \input \jobname.refs
   }
\catcode`\@=12 
\tolerance 5000
%
\let\overbar=\bar
\def\eJ{e_{\hbox{J}}}
\def\plebphi{{\varphi}}
\def\zb{{\overbar z}}
\def\yb{{\overbar y}}
\def\Db{{\overbar D}}
\def\rhob{{\overbar\rho}}
\def\dely{\partial_y}
\def\delz{\partial_z}
\def\delyb{\partial_{\overbar y}}
\def\delzb{\partial_{\overbar z}}
\def\delmu{\partial_\mu}
\def\delmub{\partial_{\overbar \mu}}
\def\delnu{\partial_\nu}
\def\delnub{\partial_{\overbar \nu}}
\def\delrho{\partial_\rho}
\def\delrhob{\partial_{\overbar \rho}}
\def\delsigma{\partial_\sigma}
\def\delsigmab{\partial_{\overbar \sigma}}

\def\dyb{d\overbar y}
\def\dzb{d\overbar z}

\def\mub{{\overbar\mu}}
\def\nub{{\overbar\nu}}
\def\rhob{{\overbar\rho}}
\def\sigmab{{\overbar\sigma}}
\def\eps{{\epsilon}}
\def\gMMb{ {g^{\mu\mub}} }

\def\inv{^{-1}}
\def\dy{dy}
\def\dz{dz}

\def\hc{{{}^{\dag}}}
\def\ab{{\overbar a}}
\def\ecab{{e_{c\ab}}}
\def\ecc{{e_{cc}}}
\def\eabab{{e_{\ab\ab}}}
\def\eaab{{e_{a\ab}}}
\def\eca{{e_{ca}}}
\def\eaa{{e_{aa}}}
\def\newecab{{\tilde e_{c\ab}}}
\def\fb{{\overbar f}}
\def\contents#1{}
\Ref\OoguriV{H. Ooguri and C. Vafa,
           \tytul{Selfduality and N=2 string magic}
          \myjournal Mod. Phys. Lett. &A5 (90) 1389;
           \tytul{Geometry of N=2 strings}
          \myjournal Nucl. Phys. B &369 (91) 469;
           \tytul{N=2 heterotic strings}
          \myjournal Nucl. Phys. B &367 (91) 83   \name{OoguriV}}
\Ref\BoniniGI{ M. Bonini, E. Gava and R. Iengo,
          \tytul{Amplitudes in the N=2 string}
          \myjournal Mod. Phys. Lett. &A6 (91) 795   \name{BoniniGI}}
\Ref\Yangi{C.N. Yang,
          \tytul{Condition for the self-duality for SU(2) gauge fields
                       on Euclidean four-dimensional space}
          \myjournal Phys. Rev. Lett. &38 (77) 1377  \name{Yangi}}
\Ref\Brihayeetal{Y. Brihaye, D.B. Fairlie, J. Nuyts and R.G. Yates,
          \tytul{Properties of the self dual equations for an SU(N) gauge
           theory}
          \myjournal J. Math. Phys. &19 (78) 2528 \name{Brihayeetal}}
\Ref\Pohlmeyer{K. Pohlmeyer,
          \tytul{On the lagrangian theory of anti self dual fields in
                   four dimensional Euclidean space}
          \myjournal Commun. Math. Phys. &72 (80) 37 \name{Pohlmeyer}}
\Ref\Parkes{Andrew Parkes,
        \tytul{On N=2 strings and classical scattering solutions of self-dual
                   Yang--Mills in (2,2) space-time}
      hepth@xxx/9110075,  ETH-TH/91-35,  Nucl. Phys. B, to be published
\contents{
Ooguri and Vafa have shown that the open N=2 string corresponds to self-dual
Yang--Mills (SDYM) and also that, in perturbation theory, it has has a
vanishing four particle scattering amplitude.  We discuss how the dynamics of
the three particle scattering implies that on shell states can only scatter if
their momenta lie in the same self-dual plane and then investigate classical
SDYM with the aim of comparing exact solutions with the tree level
perturbation theory predictions.  In particular for the gauge group SL(2,C)
with a plane wave Hirota ansatz SDYM reduces to a complicated set of algebraic
relations due to de Vega.  Here we solve these conditions and the solutions
are shown to correspond to collisions of plane wave kinks.  The main result is
that for a class of kinks the resulting phase shifts are non-zero, the
solution as a whole is not pure gauge and so the scattering seems non-trivial.
However the stress energy and Lagrangian density are confined to string like
regions in the space time and in particular are zero for the incoming/outgoing
kinks so the solution does not correspond to physical four point scattering.}
     \name{Parkes}}
\Ref\Donaldson{S.K. Donaldson,
           \myjournal Proc. London. Math. Soc. & 50 (85) 1 }
\Ref\NairS{V.P. Nair and J. Schiff,
          \tytul{A K\"ahler Chern--Simons theory and quantization of instanton
                 moduli spaces}
          \myjournal Phys. Lett. B &246 (90) 423;
\hfil\break
    K\"ahler--Chern--Simons theory and symmetries of
                    anti-self-dual gauge fields, CU-TP-521 (May 1991);
\hfil\break
  V.P.Nair, K\"ahler--Chern--Simons Theory,  hepth@xxx/9110042
\contents{
K\"ahler-Chern-Simons theory describes antiself-dual gauge fields on a four-
dimensional K\"ahler manifold. The phase space is the space of gauge
potentials, the symplectic reduction of which by the constraints of
antiself-duality leads to the moduli space of antiself-dula instantons.
We outline the theory highlighting the symmetries, their canonical
realization and some properties of the quantum wave functions. The
relationship to integrable systems via dimensional reduction is briefly
discussed.
Invited talks at Strings and Symmetries 1991, Stonybrook, May 1991 and
the XXth International Conference on Differential Geometric Methods
in Theoretical Physics, New York, June 1991 }
          \name{NairS}}
\Ref\Sokachev{S. Kalitzin and E. Sokatchev,
          \tytul{An action principle for self-dual Yang--Mills and Einstein
           equations}
          \myjournal Phys. Lett. B &262 (91) 444     \name{Sokachev}}
\Ref\MarcusOY{Neil Marcus, Yaron Oz  and Shimon Yankielowicz,
    Harmonic space, self-dual Yang--Mills and the $N=2$ string,
      hepth@xxx/9112010, TAUP-1928-91  (Dec 1991)
\contents{
The geometrical structure and the quantum properties of the recently
proposed harmonic space action describing self-dual Yang--Mills (SDYM) theory
are analyzed. The geometrical structure that is revealed is closely related
to the twistor construction of instanton solutions. The theory gets no
quantum corrections and, despite having SDYM as its classical equation of
motion, its S matrix is trivial. It is therefore NOT the theory of
the N=2 string.  We also discuss the 5-dimensional actions
that have been proposed for SDYM.}
         \name{MarcusOY}}
\Ref\ZakharovM{ V.E. Zakharov and A.V. Mikhailov,
           \tytul{Relativistically invariant two-dimensional models of field
             theory which are integrable by means of the inverse
             scattering problem method}
           \myjournal Sov. Phys. JETP & 47 (1978) 1017       \name{ZakharovM}}
\Ref\BruschiLR{M. Bruschi, D. Levi and O. Ragnisco,
        \tytul{Nonlinear partial differential equations and B\"acklund
transformations related to the 4-dimensional self-dual Yang--Mills equations}
        \myjournal Lett. Nuov. Cim. &33 (82) 263  \name{BruschiLR}}
\Ref\Ward{R.S.~Ward, The twistor approach to differential equations,
      {\it in} Quantum Gravity 2, ed. C.J.~Isham, R.~Penrose and D.W.~Sciama
      (Oxford Univ. Press 1981)}
\Ref\ColemanM{S. Coleman and  J. Mandula,
        \tytul{All possible symmetries of the S-matrix}
        \myjournal Phys. Rev. D & 159 (67) 1251  \name{ColemanM}}
\Ref\GiveonS{ Amit Giveon and Alfred Shapere,
            Gauge symmetries of the $N=2$ string,
          hepth@xxx/9203008,   IASSNS-HEP-92-14
        \contents{
We study the underlying gauge symmetry algebra of the $N=2$ string,
which is broken down to a subalgebra in any space-time background.  For
given toroidal backgrounds, the unbroken gauge symmetries
(corresponding to holomorphic and antiholomorphic worldsheet currents)
generate area-preserving diffeomorphism algebras of null
2-tori. A minimal Lie algebraic closure containing all the gauge
symmetries that arise in this way, is the background--independent
volume--preserving diffeomorphism algebra of the target Narain torus
$T^{4,4}$.  The underlying symmetries act on the ground ring of
functions on $T^{4,4}$ as derivations, much as in the case of the
$d=2$ string.  A background--independent space-time action valid for
noncompact metrics is presented, whose symmetries are
volume--preserving diffeomorphisms.  Possible extensions to $N=2$ and
$N=1$ heterotic strings are briefly discussed.}
         \name{GiveonS}}
%
\def\tytulpage{
\nopubblock\titlepage
\line{\hfil ETH-TH/92--14}
\line{\hfil March 1992}
\line{\hfil hepth@xxx/9203074}
\title{A cubic action for self-dual Yang--Mills}
\author{Andrew Parkes}
\address{Theoretische Physik,\break
ETH-H\"onggerberg,\break
8093 Z\"urich,\break
Switzerland.}
\vfil
\abstract
We make a change of field variables in the J formulation of self-dual
Yang--Mills theory.  The field equations for the resulting algebra valued
field are derivable from a simple cubic action.  The cubic interaction vertex
is different from that considered previously from the N=2 string, however,
perturbation theory with this action shows that the only non-vanishing
connected scattering amplitude is for three external particles just as for the
string.
\vfil
\endpage}
\tytulpage
\section{Introduction}

Over the years self-dual Yang--Mills (SDYM) theory has received a great deal of
attention.  The majority of this attention has focused on finding and
classifying exact solutions of the classical equations of motion, or  using
the self-dual solutions to find quantum properties of the Yang--Mills action.
For this kind of work there is no need to derive the self-duality equations
themselves from an action.  However, as realised by Ooguri and Vafa
\Rc{\OoguriV}, SDYM and self-dual gravity (SDG) in 2+2 dimensions are the
effective field theory of the N=2 string. Consistency of the string theory
implies that these self-dual systems have the remarkable property that (at
least at tree level \Rc{\OoguriV} and one-loop \Rc{\BoniniGI}) on-shell
connected scattering amplitudes vanish for four or more legs.  If we wish to
do perturbation theory of SDYM then this gives a big incentive to derive the
self-duality equations themselves from an action.

For SDG the equation of motion is the Plebanski equation which follows from a
simple cubic action.  However, for SDYM the situation is not so
straightforward.  Using the J formulation combined with an particular
parametrisation of the gauge group it is possible to write an action for SDYM
\Rc{\Yangi-\Pohlmeyer}, and indeed this action has a correctly
vanishing four point amplitude \Rc{\Parkes}.  Alternatively it is possible to
\Rc{\NairS}\ avoid the group parametrisation and obtained an action for the
J formulation by adding an extra dimension and using a generalisation of the
Wess-Zumino-Novikov-Witten (WZNW) theory \Rc{\Donaldson,\NairS}. The resulting
action is non-polynomial and whilst not an intrinsic problem this does rather
contrast with with the polynomial SDG, and is a strange reversal from the
usual properties of gravity and Yang--Mills actions.

Actions based on a very different approach were given in refs.
\Rc{\Sokachev,\MarcusOY}; harmonic space techniques were used in order to
retain the  Lorentz covariance that is lost in the J formulation.
However, in \Rc{\MarcusOY} they find that the theory exhibits no scattering
whatsoever, and so does not correspond to the N=2 string or the J formulation
which do at least have non-trivial three particle scattering.

In this paper we will show that there is a ``B\"acklund'' transformation from
the J-field of SDYM to a an algebra valued field, $\Phi$.  The main advantage
is that the $\Phi$ equation of motion follows from a simple cubic action.  We
explore tree level perturbation theory based on this action and show that
$\ge$4 leg amplitudes vanish exactly as for the N=2 string. However the cubic
vertex is not the same as that from the open N=2 string.  We then consider
Lorentz transformations and so illuminate the origin of our action; it can be
considered as arising from the J-formulation in an infinite momentum frame.
Similar methods allow us to find generalisations of the Plebanski action.  We
have explicitly checked the vanishing of the 4,5 and 6 leg amplitudes for all
the cubic actions and so can be confident they have the properties of the N=2
string.  Mostly we work in flat space and only briefly
consider the generalisation to non-trivial spaces.
\section{The cubic action}

Initially we work with analytic fields over complexified 2+2 space-time with
a complex gauge group $G_C$ and Lie algebra $g_C$.
Coordinates are denoted by $x^\alpha=(x^\mu,x^\mub)$ where $x^\mu=(y,z)$ and
$x^\mub=(\yb,\zb)$ are independent (not complex conjugates).
We take $\eps^{yz}=$ $\eps_{zy}= $ $\eps^{\zb\yb}= $ $\eps_{\yb\zb}=1$ and
$\eps_{yz\yb\zb} = \eps_{yz} \eps_{\yb\zb}$ (note a change of sign with
respect to ref. \Rc{\Parkes}).  We also take
the metric to be given by $ ds^2 =
g_{\alpha\beta} dx^\alpha dx^\beta = $ $ 2 g_{\mu\mub} dx^\mu dx^\mub = $ and
define $g = \det(g_{\mu\mub})$.
Taking a torsion free connection, and coupling constant $e$,
  the Yang--Mills field strength is
$F_{\alpha\beta}$
$=$
$[\partial_\alpha + e A_\alpha,\partial_\beta + e A_\beta]$
and  satisfies the self-duality condition
$$
F_{\alpha\beta} = + {1\over2} g \eps_{\alpha\beta\gamma\delta}
F^{\gamma\delta}
\neqn\selfdual$$
if and only if
$$
\eps^{\mub\nub} F_{\mub\nub} =0 \qquad\qquad \gMMb F_{\mu\mub}=0
\qquad\qquad \eps^{\mu\nu} F_{\mu\nu} = 0
\neqn\sdcondits
$$
 We now restrict to  trivial topology and a flat metric given by
$\eta_{\mu\mub} dx^\mu dx^\mub = $ $ \dy\; \dyb - \dz\; \dzb$.  Then,
\E\sdcondits\ implies that there exist $D\in G_C$ and $\Db\in G_C$ such that
$e A_\mu = D^{-1} \delmu D$ and $e A_\mub = \Db^{-1}\delmub \Db$.   Defining
$J = D \Db^{-1}$ reduces the duality conditions to the following well known
field equation for J
$$
\eta^{\mu\mub} \; \delmub ( J^{-1} \delmu J) =0
\neqn\Jeqn$$
At this point we are motivated by the similarity to the chiral model in 1+1
space-time and its reformulation as discussed for example in ref.
\Rc{\ZakharovM}.
Defining $B_\mu$ by
$$
e B_\mu = J\inv \delmu J
\neqn\Bdefn$$
we see that given a solution of \E\Jeqn\ then, generically, there exists
some  $\Phi\in g_C$ such that
$$
B_y =  \delzb \Phi  \qquad B_z  = \delyb \Phi
\neqn\Bphi$$
This automatically solves \E\Jeqn.
Finally \E\Bdefn\ implies  $[\dely + e B_y , \delz + e
B_z] =0$ which yields the following field equation for  $\Phi$
$$
\eta^{\mu\mub} \delmu \delmub \Phi
                      + e \eps^{\mub\nub} \delmub \Phi \; \delnub\Phi =0
\neqn\Phieom
$$
This equation appeared in ref. \Rc{\BruschiLR} as part of more general
study of B\"acklund transformations in SDYM.  For us the point of this
construction is that \E\Phieom\ is the Euler-Lagrange equation of the
following action
$$
S={1\over C_R}\hbox{Tr} \int d^4x
\left[-{1\over2} \eta^{\mu\mub} \; \delmu\Phi \;  \delmub\Phi
     + {e\over3} \eps^{\mub\nub} \; \delmub\Phi \; \delnub \Phi \; \Phi \right]
\neqn\Phiaction$$
where as usual $\Phi = \Phi^a T_a$, $[T_a,T_b] = f_{abc}g^{cd}T_d$ and
$\hbox{Tr} (T_a T_b) = - C_R g_{ab}$.

In the above we have used the J formulation only in order to exhibit the gauge
independence of the construction.  A more direct route is to make the
Yang--Mills gauge choice $A_\mub=0$ in which case $B_\mu = A_\mu$ and then
\E\Bphi\ follows as the general solution of $\eta^{\mu\mub} F_{\mu\mub} = 0$.
Finally $F_{yz}=0$ gives \E\Phieom\ as the field equation for $\Phi$.

Conversely, given a field $\Phi$ then we use \E\Bphi\ to define fields $B_\mu$
which  automatically satisfy the ``conservation law''
$$
\eta^{\mu\mub} \delmub B_\mu = 0
\neqn\divB$$
The field equation for $\Phi$ then implies we can find a $J\in G_C$ such that
\E\Bdefn\ holds true, and so we recover \E\Jeqn.  In particular, $A_\mu=B_\mu$
with $A_\mub=0$ gives a self-dual Yang--Mills field strength.

We have seen that \E\Bphi\ maps solutions of \E\Jeqn\ to solutions of
\E\Phieom\ and vice versa.  It is also a B\"acklund transformation between two
fields in the sense that consistency of the transformation implies that both
fields must satisfy their equations of motion.  Thus, we only have shown
equivalence of the on-shell theories; however the viewpoint in this paper
shall be to investigate \E\Phiaction\ for both on and off-shell fields.

In order to have four real dimensions and  a real gauge group we must impose
some reality conditions.
The standard choice is to take $(x^\mu)^* = x^\mub$, $(A_\mu)\hc = - A_\mub$
and $J\hc=J$, but in this case the B\"acklund transformation is generally not
compatible with a simple reality condition on $\Phi$.
It is much simpler to implement the reality condition by $(x^\alpha)^* =
x^\alpha$ and $A_\alpha{}\hc = - A_\alpha$. In this case $J\in G_R$ and
$\Phi\hc = - \Phi \in g_R$.  Notice that if we want a real gauge group then we
are forced to take a 2+2 space-time; this method cannot treat SU(2)
instantons in Euclidean space.

\section{ Perturbation theory}

A striking property of the N=2 string and so of these self-dual systems is that
in tree level perturbation theory the on shell connected Feynman diagrams sum
to zero for four or more external legs.  For consistency it should be that the
perturbation theory based on \E\Phiaction\ should have the same property.

It is very convenient to define
$$
a_{ij} = \eps^{\mu\nu} k_{\mu i} k_{\nu j}  \qquad
\ab_{ij} = \eps^{\mub\nub} k_{\mub i} k_{\nub j}  \qquad
\neqn\aabdefns$$
$$
k_{ij} = \eta^{\mu\nub} k_{\mu i} k_{\nub j}  \qquad
s_{ij} = k_{ij} + k_{ji} \qquad
c_{ij} = k_{ij} - k_{ji}
\neqn\kscdefns$$
{}From the action \E\Phiaction\ we have the usual propagator $<\!\Phi^a(k)
\Phi^b(-k)\!> \;\propto\;  g^{ab}/k^2$ whilst our off-shell
three leg vertex is simply $<\! \Phi^a(k_1)
\Phi^b(k_2) \Phi^c(-k_1 - k_2) \! > \; \propto \; e \; \ab_{12} f_{abc}$.
 We emphasise that the three leg on-shell ampliude is not identically zero
and so we do not have the same theory as in ref. \Rc{\MarcusOY}.
or comparison the N=2 string gives a vertex proportional to $c_{12}$ instead
of $\ab_{12}$.
We can  generate many identities by taking the trivial identity
$$
k_{\mu i} k_{\nu j} k_{\rho m} + \hbox{antisym(ijm)} \equiv 0
\neqn\kkkiden$$
and  contracting the free space-time indices with various tensors.   For
example, we find
$$
\ab_{ij} k_{mn} + \hbox{cycl(ijn)} \equiv 0
\neqn\abkiden$$
which leads to
$$
k_{ii} = \Sigma_1^4 k_i = 0
\qquad\Longrightarrow\qquad
{\ab_{12} \ab_{34} \over s_{12}} + {\ab_{13} \ab_{24} \over s_{13}}  \equiv 0
\neqn\ababiden$$
This can easily be used to show the vanishing of the on-shell tree-level
four leg
amplitude without the need for a four leg vertex. The corresponding ``cc/s''
identities have extra terms which signify the need for a four leg vertex, as
in perturbation theory based directly on the J formulation.

Above 4 legs I found it much better not to try to derive more identities but
simply to apply the much more inelegant brute-force (and ideal for computer
algebra) technique of solving for the components of momenta so as to put all
external legs on shell.  In this way I have been able to verify that the action
\E\Phiaction\ leads to vanishing of the on-shell tree-level  5 and 6 leg
diagrams.
It is important to go to at least 6 legs because only then do we fully probe
the off-shell structure of the vertex because with 4 and 5 external legs all
diagrams are such that at least one leg of every vertex is on shell.  For the
same reason I expect that if the 4,5, and 6 leg amplitudes vanish then we can
be confident that all tree-level  amplitudes with more legs will also vanish.

At  one loop loop it is easy to see that the tadpole and propagator
corrections vanish both on- and off-shell.  The three point amplitude is
propartio9nal to $\ab_{12}{}^3$ and is both UV and IR divergent.  Thus, as
expected the theory is not renormalisable.  In the string version the UV
divergence is removed by the modular invariance leaving just IR divergences
(as pointed out in \Rc{\OoguriV,\BoniniGI}) and it is interesting to
speculate that some kind of ``soft-photon'' sum could remove all these as
well.

\section{Reduction to two dimensions}

Having seen that the action \E\Phiaction\ has the expected properties in tree
level perturbation theory it is natural to  further explore its properties by
looking at two dimensional version.

Suppose we trivially reduce by forcing fields to depend only on $\eta_i =
k_{\mu i} x^\mu + k_{\mub i} x^{\mub}$ with $i$=1,2  and $k_{11}=k_{22}=0$.
Then the the J equation \E\Jeqn\ becomes
$$
k_{12} \partial_1 ( J\inv \partial_2 J) + k_{21} \partial_2 ( J\inv \partial_1
J) = 0
\neqn\cmwz$$
which is simply the equation of motion of the principal chiral model with a
Wess-Zumino term proportional to $c_{12}$.  It becomes a WZNW
theory at $k_{12}=0$ or $k_{21} = 0$.

The B\"acklund transformation \E\Bdefn,\E\Bphi\ reduces to
$$
\pmatrix{  k_{y1} & k_{y2} \cr
           k_{z1} & k_{z2} \cr}
\pmatrix{ J\inv \partial_1 J \cr
          J\inv \partial_2 J \cr}
=
\pmatrix{  k_{\zb1} & k_{\zb2} \cr
           k_{\yb1} & k_{\yb2} \cr}
\pmatrix{  \partial_1 \Phi \cr
           \partial_2 \Phi \cr}
\neqn\Jphiintwo$$
The two matrices involved are both non-singular only if $a_{12} \ab_{12} \neq
0$ and from the identity $ a_{12} \ab_{12} = k_{11} k_{22} - k_{12} k_{21}$ we
see this needs $k_{12} k_{21} \neq0$. Thus, the transformation fails precisely
in the case in which the model is a WZNW model.  However, away from this
singular point we can proceed to do the transformation.  In this case
\E\Phieom\
reduces to
$$
k_{12} \partial_1 \partial_2 \Phi + e \ab_{12} [\partial_1 \Phi, \partial_2
\Phi] = 0
\neqn\Phieomintwo$$
coming from the reduced action
$$
S={1\over C_R}\hbox{Tr} \int d^2\eta
\left[-{s_{12}\over2} \partial_1 \Phi \; \partial_2 \Phi
     - {e \ab_{12}  \over3}  [\partial_1 \Phi ,\partial_2 \Phi ] \; \Phi
\right]
\neqn\Phiactionintwo$$
This provides an action for the the chiral model with Wess-Zumino term in 1+1
dimensions that does not require any explicit parametrisation of the group nor
an extension into 3 dimensions.  The action differs from the one without
Wess-Zumino term only in a rescaling of the coupling constant.  It may also be
a disadvantage that the transition to algebra rather than group valued fields
loses a lot of the group structure.  The transition itself is only a local
construction.  Despite this it would be interesting to look at the 1+1
dimensional quantum theory based on \E\Phiactionintwo.

Alternatively we  can solve \E\cmwz\  by writing
$$
J\inv \partial_1 J = k_{12} \partial_1 P
\qquad
J\inv \partial_2 J = -  k_{21} \partial_2 P
\neqn\cmwzback$$
and the identical construction leads to the action
$$
S={1\over C_R}\hbox{Tr} \int d^2\eta
\left[-{s_{12}\over2}a\partial_1 P \partial_2 P
     - {k_{12} k_{21} \over3}  [\partial_1 P ,\partial_2 P] P
\right]
\neqn\Phiaction$$
Ignoring the fact that \E\cmwzback\  is singular than the theory is clearly
classically
trival at the WZNW case.

\section{Lorentz symmetry}

Our starting point was the self-duality equation, \E\selfdual, which posseses
a manifest SO(2,2) Lorentz symmetry.  However, both the J formulation and the
formulation given above obscure this symmetry by ``solving'' \E\selfdual\ in a
non-covariant fashion.  In particular, we chose a symplectic structure when we
made a split into $x^\mu$ and $x^\mub$ and did not treat them equally.  The
remaining manifest space-time symmetry of \E\Phiaction\ with real coordinates
is a meagre SL(2,R) (in comparison, the J formulation in a C$^{1,1}$ space has
the more inviting U(1,1) symmetry).  Despite this noncovariance it is still
instructive to consider the effects of an SO(2,2) Lorentz transformation on
\E\Phiaction\ and so vary the symplectic structure within SO(2,2)/SL(2,R).  We
consider
$$
\eqalign{
\delmu \to & f_1 \delmu - f_2 \eps_{\mu\nu} \eta^{\nu\nub} \delnub \cr
\delmub \to & \fb_1 \delmub - \fb_2 \eps_{\mub\nub} \eta^{\nu\nub} \delnu \cr
}\neqn\lorentz$$
where $f_i$ and $\fb_i$ are constant parameters subject to the restriction
$ (f_1 \fb_1 + f_2 \fb_2) =1$
in order  that $\eta^{\mu\mub} \delmu\delmub $ is invariant.

Under such a transformation the  action \E\Phiaction\ is not
invariant but becomes
$$
\eqalign{
S= & {1\over C_R}\hbox{Tr} \int d^4x  \bigg[
   -{1\over2} \eta^{\mu\mub} \; \delmu\Phi \;  \delmub\Phi \cr
   &  + {e\over3}\left(
\fb_1^2 \eps^{\mub\nub} \; \delmub\Phi \; \delnub \Phi \; \Phi
+ \fb_1 \fb_2 \eta^{\mu\mub}  [\delmu \Phi , \delmub \Phi] \; \Phi
+ \fb_2^2 \eps^{\mu\nu} \; \delmu\Phi \; \delnu \Phi \; \Phi \right) \bigg]
\cr}
\neqn\boostedphiaction$$
and the momentum part of the corresponding three point vertex is  $\fb_1^2
\ab_{12} + \fb_1 \fb_2 c_{12} + \fb_2^2 a_{12}$ Again, explicit
calculations show that the tree-level  4,5 and 6 point
 scattering amplitudes vanish.
This is
not surprising since we have only simply done a global space-time boost.

It is also possible to use this procedure to generate ``new'' actions for
self-dual gravity which have the same properties for the scattering
amplitudes.
Consider the generalised Plebanski action $S=\int d^4x L$ with
$$
\eqalign{
L = &    {1\over2} \eta^{\mu\mub} \; \delmu\plebphi \;  \delmub\plebphi \cr
& + \eaa  \;  \eps^{\mu\nu} \eps^{\rho\sigma}
             \delmu \delrho \plebphi \; \delnu \delsigma\plebphi \; \plebphi
 + 2 \eca \;  \eps^{\mu\nu} \eta^{\rho\sigmab}
             \delmu \delrho \plebphi \; \delnu \delsigmab \plebphi \; \plebphi
\cr
& + 2 \eaab \;  \eps^{\mu\nu} \eps^{\mub\nub}
             \delmu \delmub \plebphi \; \delnu\delnub \plebphi \; \plebphi  \cr
& + 2 \ecc  \;  \eta^{\mu\nub} \eta^{\rho\sigmab}
         ( \delmu \delrho \plebphi \; \delnub \delsigmab \plebphi \; \plebphi
            - \delmu \delsigmab \plebphi \;  \delnub \delrho \plebphi \;
\plebphi ) \cr
& +  2 \ecab \;  \eps^{\mub\nub} \eta^{\rho\sigmab}
            \delmub \delrho \plebphi \; \delnub \delsigmab \plebphi  \;
\plebphi \cr
& +  \eabab \;  \eps^{\mub\nub} \eps^{\rhob\sigmab}
            \delmub \delrhob \plebphi \; \delnub \delsigmab \plebphi \;
\plebphi \cr
}\neqn\genplebaction$$
The usual Plebanski equation corresponds to the case where only $\eaab$ is
non-zero.
This  gives rise to a three point  vertex proportional to
$$
\eaa    a_{12}^2 +
\eca   c_{12} a_{12} +
\eaab    a_{12} \ab_{12} +
\ecc    c_{12}^2 +
\ecab    c_{12} \ab_{12} +
\eabab   \ab_{12}^2
\neqn\genplebvert$$
and by explicit calculation I find that the on-shell 4 and 5 point scattering
amplitudes vanish if
$$
\eqalign{
\eaa \eabab  - \eca \ecab + \eaab \ecc + \ecc^2 & \equiv 0  \cr
 ( \ecab^2 - 4 \ecc \eabab)
(2 \eaab \eabab - \ecab^2  - ( \ecab^2 - 4 \ecc \eabab))^2
 -\qquad  & \cr
(4 \eabab^2 \eca - 2 \eaab \eabab \ecab +  2 \ecab ( \ecab^2 - 4 \ecc
\eabab))^2
& \equiv 0 \cr}
\neqn\genplebcondit$$
I write the conditions in this form because in this way they are most
convenient for the computer calgebra calculations.  I found it best to solve
the first condition for $\eaa$ and then to introduce $\newecab$ by setting
$\ecc = (\ecab^2 - \newecab^2)/(4 \eabab)$ and solving the second condition
for $\eca$.  In this fashion I have managed to do the 6 point function with
random numeric values for the on-shell incoming momenta and find the graphs
always sum to zero.  So with the conditions \E\genplebcondit\ the tree-level
6 particle
amplitude also vanishes on-shell.

In fact this solution arises simply from applying the transformation
\E\lorentz\ to the original Plebanski action; we get
$$
\eqalign{
\eaa   =&   f_1^2 \fb_2^2 \qquad\qquad
\eca   =   \fb_1 \fb_2 f_1^2 -  \fb_2^2 f_1 f_2 \qquad\qquad
\ecc   = - \fb_1 \fb_2 f_1 f_2  \qquad\qquad \cr
\eaab  =&  \fb_1^2 f_1^2 + \fb_2^2 f_2^2 \qquad\qquad
\ecab  = - \fb_1^2 f_1 f_2 +  \fb_1 \fb_2 f_2^2 \qquad\qquad
\eabab =   \fb_1^2 f_2^2 \cr
}\neqn\esoln
$$

Lorentz boosts change the form of the action but not the
property that scattering amplitudes vanish for more than three legs.
Thus, it is natural to ask what would happen if we did such boosts starting
from the J formulation rather that \E\Phiaction.
For simplicity,  consider the special case that $f_1=\fb_1=1$,
$\fb_2=0$ and  $f_2$ is arbitrary.  Then in the J formulation perturbation
theory
 the
cubic vertex is $\eJ  c_{12} f_{abc} $ which gets boosted to
$$
 \eJ ( f_2^2 \ab_{12} + c_{12}) f_{abc}
$$
Hence the interaction term of \E\Phiaction\ is naturally generated from the J
formulation.  Furthermore, suppose we go to an ``infinite momentum frame'' by
taking $f_2 \to \infty$, but simultaneously taking $\eJ \to 0$ in such a way
that $\eJ f_2^2 = e$ remains finite (and small).  Then in this limit  the $ \eJ
c_{12}$ cubic vertex vertex  disappears. The J formulation also has
vertices with more legs but these necessarily come with higher powers of $\eJ$
and so also vanish in the above limit.  The only remaining vertex is the $e
\ab_{12}$
vertex that one would get from \E\Phiaction.  Thus we can interpret
\E\Phiaction\ as arising from the J formulation in an ``infinite momentum
frame'', and in the process we have reduced a non-polynomial action to a cubic
action.

Bye the same process we can convert the usual action for the Plebanski equation
(which is \E\genplebaction\ with only $\eaab$ non-zero) to an action in which
only $\eabab$ is nonzero.  Such an action solves \E\genplebcondit\ and so has
vanishing scattering amplitudes for 4 or more legs.

Finally, consider a boost of the cubic action \E\Phiaction\ with the
parameters $f_1 \fb_1=1$, $f_2 = \fb_2=0$ and $\fb_1$ arbitrary.  It is easy
to see that the only effect that this has is the rescaling of the coupling
constant
$$
e \to e \fb_1^2
\neqn\erescaling$$
This means that if we have a solution to \E\Phieom\ at coupling $e$ then we
can generate another solution but at coupling $e \fb_1^2$ by means of such a
Lorentz boost.  However, if we now restrict our attention only to Lorentz and
gauge invariant features of the theory, one might expect that we can than undo
the Lorentz boost without changing the coupling.  Thus Lorentz invariant
sacattering features seem to be the same for all couplings, and so the same as
in the linearised theory.  Matching the often repeated statement that there is
no scattering on SDYM \Rc{\Ward}.  Clearly there is scattering, but possibly
not of
any Lorentz invariant quantity.  Indeed ``correcting'' the loss of manifest
Lorentz invariance by use of harmonic space techniques also ended up with
vanishing scattering amplitudes \Rc{\MarcusOY}.

The N=2 string vertex is definitely not Lorentz invariant; it could well be
that any attempt to enforce such Lorentz symmetry is doomed to bring along
with it triviality.  We also note that SDYM is an integrable theory without a
trivial S-matrix and so has somehow avoided the Coleman-Mandula theorem
\Rc{\ColemanM}.  There are two obvious escape routes; the theorem assumes
Lorentz invariance and that there is non-trivial scattering for most incoming
momenta.  The N=2 string evades on both these counts.  Attempting to
re-instate Lorentz invariance is difficult because of the scarcity of
candidate Lorentz invariant momentum dependent on-shell three point functions.

\section{Generalisation to curved space}

The previous considerations for SDYM were for a flat background space-time. In
this section we do the same things but starting with an background metric
$g_{\mu\mub}$ which satisfies the ``K\"ahler'' conditions
$$
\eps^{\mu\nu} \delmu g_{\nu\nub} =
\eps^{\mub\nub} \delmub g_{\nu\nub} = 0
\neqn\symplectic$$
We assume that it is possible to go to the gauge $A_\mub = 0$ then
$g^{\mu\mub} F_{\mu\mub}=0$ becomes
$$
g^{\mu\mub} \delmub A_\mu =0
\neqn\divA$$
which is solved by
$$
A_\mu = \eps_{\mu\nu} g^{\nu\nub} \delnub \left( \sqrt{-g} \Phi \right)
\neqn\Amu$$
where we have used $ \delmub( g g^{\mu\mub})=0$ and the
 $1/\sqrt{-g}$ is inserted so that $\Phi$ transforms as a scalar
(rather than scalar density).

The remaining condition $F_{yz}=0$ then gives
$$
\delmu \left[ g^{\mu\mub} \delmub\left( \sqrt{-g} \Phi \right) \right]
 -{e\over g} \eps^{\mub\nub}
\delmub \left( \sqrt{-g} \Phi \right) \;
\delnub \left( \sqrt{-g} \Phi \right)  =0
\neqn\bentPhieom$$

In order to be able to derive this from an action we impose conditions on the
background metric.  In particular we take $\partial_\alpha g =0$.  This is a
reasonable condition to take because it makes the Riemann curvature tensor of
the background self-dual and explained in ref. \Rc{\OoguriV} it gives rise to
the Plebanski equation and cubic action discussed earlier.  When
$g=\det(g_{\mu\mub})$ is constant then \E\bentPhieom\ follows from the action
$$
S={1\over C_R}\hbox{Tr} \int d^4x \; (-g)
\left[+{1\over2}  g^{\mu\mub} \; \delmu\Phi \;  \delmub\Phi
     +  {e\over3 \sqrt{-g}} \eps^{\mub\nub} \;
              \delmub\Phi \; \delnub \Phi \; \Phi \right]
\neqn\bentPhiaction$$
which reduces to \E\Phiaction\ in flat space.

Of course the above construction heavily uses the fact that our space-time is
topologically trivial and requires more care in non-trivial topologies.

\section{Conclusions}

We have shown how to describe SDYM by means of a cubic action, and that the
resulting action gives scattering amplitudes with similar properties to the
N=2 string.  It would certainly be interesting to explore the quantum theory
of this action further, despite the fact that naive power counting suggests it
is  nonrenormalizable.

Finally we remark that the action for SDG of Giveon and Shapere \Rc{\GiveonS}
has similarities to \E\bentPhiaction\ and it would certainly be interesting to
couple them together.  Presumably this would correspond to unravelling the
geometry of the  heterotic $N=2$ string discussed in ref.  \Rc{\OoguriV}.

\ack
{I am grateful to J. Schiff for drawing my attention to ref. \Rc{\BruschiLR}.
I  thank the ETH for hospitality during my fellowship.  I am
also grateful to the people of Great Britain and Switzerland for their
financial support through the Royal Society of Great Britain and the Swiss
National Science Foundation, respectively.}
\refout
\bye